\documentclass[journal,10pt]{IEEEtran}
\usepackage[T1]{fontenc}
\usepackage{times}
\usepackage{color}
\usepackage{amsmath}
\usepackage{amssymb}
\usepackage{verbatim}
\usepackage{stmaryrd}
\usepackage{stackrel}
\usepackage{graphicx}
\usepackage{cite}
\usepackage{cases}
\usepackage{caption}
\usepackage{stackrel}
\usepackage{multicol}
\usepackage[USenglish]{babel}
\usepackage{array}
\usepackage{multirow}
\usepackage{amsfonts}
\usepackage{float}
\usepackage{balance}
\usepackage{ragged2e}
\usepackage{subfigure}
\usepackage{physics}
\usepackage[ruled,linesnumbered]{algorithm2e}
\usepackage{stfloats}
\usepackage{tikz}
\usepackage{autobreak}

\renewcommand{\raggedright}{\leftskip=0pt \rightskip=0pt plus 0cm}
\raggedright

\newtheorem{remrk}{\textbf{\textit{Remark}}}

\newtheorem{prop}{\textbf{\textit{Proposition}}}
\allowdisplaybreaks

\begin{document}
\title{Intelligent Trajectory Planning in UAV-mounted Wireless Networks: A Quantum-Inspired Reinforcement Learning Perspective}
\author{Yuanjian~Li, A. Hamid~Aghvami~\IEEEmembership{Fellow,~IEEE} and Daoyi Dong~\IEEEmembership{Senior Member,~IEEE}\thanks{Yuanjian Li and A. Hamid Aghvami are with Centre for Telecommunications Research (CTR), King's College London, London WC2R 2LS, U.K. (e-mail: yuanjian.li@kcl.ac.uk; hamid.aghvami@kcl.ac.uk).}
\thanks{Daoyi Dong is with the School of Engineering and Information Technology,
University of New South Wales, Canberra ACT 2600, Australia. (e-mail:
d.dong@adfa.edu.au).}
\thanks{This work has been submitted to the IEEE for possible publication. Copyright may be transferred without notice, after which this version may no longer be accessible.}}

\maketitle

\thispagestyle{empty}
\begin{abstract}
In this paper, we consider a wireless uplink transmission scenario in which an unmanned aerial vehicle (UAV) serves as an aerial base station collecting data from ground users. To optimize the expected sum uplink transmit rate without any prior knowledge of ground users (e.g., locations, channel state information and transmit power), the trajectory planning problem is optimized via the quantum-inspired reinforcement learning (QiRL) approach. Specifically, the QiRL method adopts novel probabilistic action selection policy and new reinforcement strategy, which are inspired by the collapse phenomenon and amplitude amplification in quantum computation theory, respectively. Numerical results demonstrate that the proposed QiRL solution can offer natural balancing between exploration and exploitation via ranking collapse probabilities of possible actions, compared to the traditional reinforcement learning approaches which are highly dependent on tuned exploration parameters.
\end{abstract}


\section{Introduction}

\IEEEPARstart{U}{nmanned} aerial vehicle (UAV) has been recognised as a promising technique to facilitate wireless communications in recent years, due to its attractive advancements such as flexible mobility, on-demand deployment and cost effectiveness \cite{zeng2019accessing, wang2017taking}. Compared to terrestrial wireless communication scenarios, one of the most notable features of UAV-mounted wireless networks is the controllable adjustments of UAV's flying trajectory, which can offer favourable wireless channel quality \cite{hu2020reinforcement}. This feature encourages the concern of UAV's trajectory design, which is a key research objective in UAV-aided networks. To solve optimal trajectory planning problem of UAV-based networks, reinforcement learning (RL) has been leveraged, for its ability to learn in a "trial-and-error" manner without explicit knowledge of the environment \cite{yin2019intelligent, challita2019interference}. 

Balancing exploration and exploitation remains the inherent challenge of RL-based intelligent systems, which poses significant impacts on learning efficiency and quality, e.g., $\epsilon$-greedy and Boltzmann action selection strategies \cite{dong2008quantum, wang2020thirty, li2020quantum}. On one hand, $\epsilon$-greedy method renders that a random action is executed with probability $\epsilon\in[0,1]$, and the optimal action is selected with probability ($1 - \epsilon$) according to the developed action selection policy. This method is simple and effective. However, one of its drawbacks is that it selects uniformly among all possible actions while exploring, which means that it cannot distinguish the next-to-optimal action from other possible counterparts. 
 On the other hand, Boltzmann (or the Softmax) exploration method introduces an action selection probability $\exp(Q(s,a)/\tau)/(\sum_{i}\exp(Q(s,a^{i})/\tau))$ based on the Q function $Q(s,a)$ of state $s$ and action $a$, where the parameter $\tau$ represents the \textit{temperature} in the Boltzmann distribution. 
 However, finding a good $\tau$ which can properly balance exploration and exploitation is difficult. The parameters $\epsilon$ and $\tau$ pose significant impacts on the convergence performance and the quality of learning output, which makes it necessary to develop new action selection strategy for RL.

Recently, with the advancement of quantum computation techniques, it is  believed to be a promising direction to adopt quantum mechanism into the field of machine learning \cite{biamonte2017quantum}. Dong \textit{et al.} \cite{dong2008quantum} proposed the concept of quantum reinforcement learning (QRL), in which QRL was applied to solve the typical grid-world problem. 
Thereafter, in \cite{dong2010robust}, Dong \textit{et al.} introduced quantum-inspired reinforcement learning (QiRL) into the field of navigation control of autonomous mobile robots. 
Fakhari \textit{et al.} \cite{fakhari2013quantum} applied QiRL approach into unknown probabilistic environment, in which the robustness of QiRL solution was demonstrated.
Li \textit{et al.} \cite{li2020quantum} compared QRL with several conventional RL (CRL) models\footnote{The abbreviation "CRL" denotes the RL methods without involving neural networks, distinguishing itself from deep reinforcement learning (DRL).} in human decision-making scenarios, suggesting that value-based decision-making can be illustrated by QRL at both the behavioural and neural levels. However, QiRL is now still in its infancy, and has not been yet introduced into the field of UAV-aided networks, e.g., solving path planning problem for UAV.

In this paper, a novel RL algorithm inspired by quantum mechanism, which is independent  on exploration parameters, is applied to tackle the trajectory planning problem in UAV-aided uplink transmission scenario. Specifically, in this proposed QiRL solution, balancing exploration and exploitation is realized in a manner inspired by the collapse phenomenon of quantum superposition and the quantum amplitude amplification.\footnote{In QRL, it is expected to implement real quantum computation on practical quantum computers, while QiRL algorithm invokes several ideas from quantum theory and is still in the frame of CRL which can be directly conducted on traditional computers.}
Different from \cite{dong2008quantum} and \cite{dong2010robust}, we extend the quantum explanation of QiRL from fixed rotation angles to their flexible counterparts, which is an alternative of \cite{li2020quantum} and \cite{fakhari2013quantum}. Besides, we also relax the limitation of linear function mapping in \cite{li2020quantum} and that of empirical rotation angle setting in \cite{fakhari2013quantum}. We aim at providing the first exploration of emerging QiRL for UAV-aided wireless networks.

\section{System model}
This work concentrates on the uplink transmission scenario consisting of one UAV$\footnote{Without loss of generality, we focus on the system model with one single UAV, while the proposed QiRL algorithm can be similarly applied to other UAVs. The multi-UAV scenario is of importance to be evaluated, which is out of the scope of this paper and left as one of future research directions. }$ and $K$ ground users (GUs), e.g., Fig. 1, in which the location of each ground user is denoted as $\vec{D}_{k} = (x_{k},y_{k},0)$ where $k\in\{1,2,\dots,K\}$. It is assumed that all the GUs are uploading their messages in a  frequency division multiplexing manner. Thus, each GU transmits sorely on its assigned channel and inner-channel interference  can be approximately ignored. Besides, the UAV is assumed to fly with constant velocity $V$ (m/s) and fixed altitude $H$ (m).$\footnote{The UAV's altitude $H$ is assumed as a fixed parameter, which may correspond to the lowest altitude required for terrain or building avoidance, under the regulation of local laws in practice.}$ A practical assumption on the availability of network information  is applied, in which the UAV cannot obtain any environment knowledge, e.g., transmit power of the GUs, locations of the GUs, and can only observe the received signals from the GUs. The goal of the UAV is to maximize the expected sum uplink transmit rate (ESUTR) of the GUs via intelligently adjusting its flying trajectory from the start location $\vec{L}_{0} = (x_{0},y_{0},H)$ to the destination $\vec{L}_{F} = (x_{F},y_{F},H)$. Assume that the feasible region where the UAV can explore is a rectangular area $\left[x_{0}, x_{F}\right]\times\left[y_{0}, y_{F}\right]$, denoted as $\Phi$ for clarity. To make the trajectory design tractable, the entire trajectory is discretized into $F$ equal-spacing steps, via evenly quantifying the time horizon into $F$ time slots, where the length of each time slot is predefined as $T$ (s). Furthermore, the 3-dimensional Cartesian coordinate at the beginning of each time slot can be given by $\mathcal{L} = \{\vec{L}_{0},\vec{L}_{1},\dots,\vec{L}_{F}\}$, in which $\vec{L}_{0}\preceq\vec{L}_{f}\preceq\vec{L}_{F}, \forall f\in\left[0, F\right]$, where $\preceq$ represents element-wise inequality. 
\begin{figure}[tbh]
	\centering
	\includegraphics[width=0.3\textwidth]{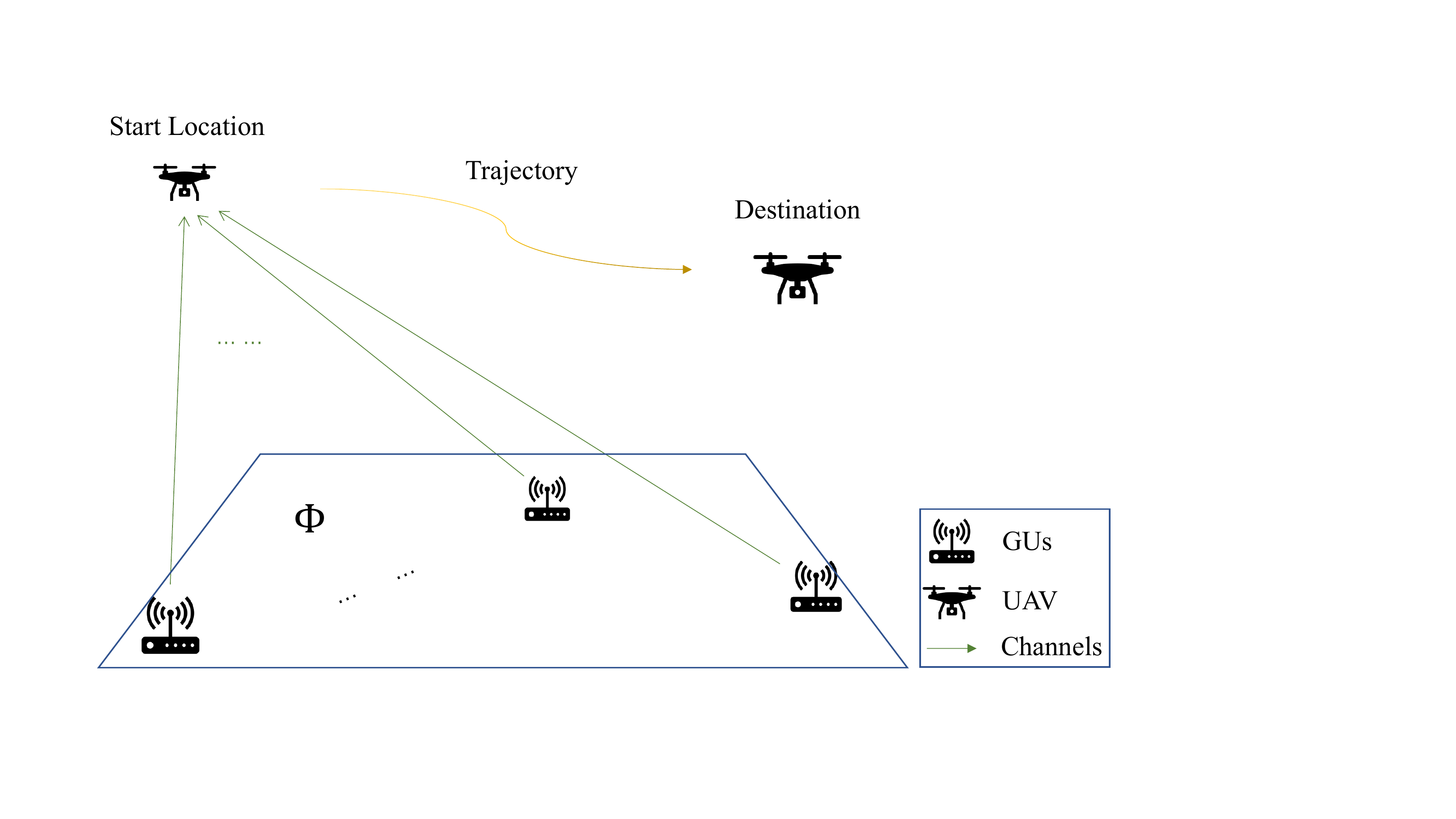}
	\caption{System Model}
	\label{System Model}
	\vspace{-.2cm}
\end{figure}

The large-scale path loss model on the sub-6 GHz band is considered to characterize the channel gains for wireless links between the UAV and all GUs \cite{challita2019interference}, which can be given by
\begin{equation}
	PL_{fk}(\text{dB}) = 20\lg(d_{fk}) + 20\lg(\varphi) - 147.55,
\end{equation}
where $d_{fk}=\Vert{\vec{L}}_{f} - {\vec{D}}_{k}\Vert$ denotes the Euclidean distance between the UAV at sampled location ${\vec{L}}_{f}$ and the GU $k$, and $\varphi$ represents the carrier frequency. Note that we herein take line-of-sight (LoS)-dominated channel gain as an example to evaluate the proposed system model, which is suitable for suburban or rural scenario, i.e., the channel gain between the drone and GUs can be characterized by the distance-based fading channel model.\footnote{This work focuses on strong LoS path loss channel model and the effects of small-scale fading (e.g., Rician fading or Nakagami-$m$ fading) is omitted. Besides, non-line-of-sight (NLoS) channel gain can also be easily integrated into the proposed model via involving extra NLoS fading component, which means the proposed algorithm is still applicable for NLoS case and this case is omitted for conciseness.}

The received signal-to-noise ratio (SNR) at the UAV from the  GU $k$ can be derived as 
\begin{equation}
		\Gamma_{fk} = \frac{P_{k}}{\sigma_{k}^{2} 10^{\frac{PL_{fk}}{10}}},
\end{equation}
where $P_{k}$ represents the uplink transmit power of the GU $k$ and $\sigma_{k}^{2}$ denotes the power of additive white Gaussian noise.

\section{Problem Formulation}
In this paper, we focus on maximizing the ESUTR for the UAV travelling from the predefined start location to the destination, via finding its optimal trajectory. It is straightforward to conclude that, at each sampled UAV coordinate $\vec{L}_{f}$, the sum uplink transmission rate can be characterized by $\sum_{k=1}^{K}\omega_{k}\log(1+\Gamma_{fk})$ where $\omega_{k}$ means the bandwidth occupied by the GU $k$. 
Furthermore, the problem of ESUTR maximization can be stated as
\begin{subequations}
\begin{align}
{\max\limits_{\mathcal{L}}}\frac{1}{F}&\sum_{f=1}^{F}\sum_{k=1}^{K}\omega_{k}\log(1+\Gamma_{fk}),\label{ProposedProblem}\\
&\text{s.t.}\hspace{.15cm}\Vert{\vec{L}}_{f} - {\vec{L}}_{f-1}\Vert = VT, \label{Constraint1} \\ 
&\hspace{.55cm}\vec{L}_{0}\preceq\vec{L}_{f}\preceq\vec{L}_{F},\label{Constraint2} \\
&\hspace{.55cm}FT\leq E,\label{Constraint3} \\
&\hspace{.55cm}\sum_{k}\omega_{k}\leq B, \label{Constraint4}\vspace{-.4cm}
\end{align}
\end{subequations}
where $B$ indicates bandwidth capacity of the system and $ E$ represents the maximum flight time threshold. Note that the constraint (\ref{Constraint1}) ensures that the flying distance between arbitrary adjacent time slots is fixed as the UAV's roaming capacity $VT$,  the constraint (\ref{Constraint2}) makes sure that the UAV's trajectory is exclusively within the feasible regime, the constraint (\ref{Constraint3}) declares that the maximum exploration time $FT$ is constrained by the on-board power capacity of the UAV and the constraint (\ref{Constraint4}) limits that the sum of each GU's occupied bandwidth should lie in the range of available bandwidth resource. 

The proposed problem (3) cannot be tackled via traditional optimization approaches due to the lack of environment information but can be solved by model-free RL algorithms in a "trial-and-error" manner, e.g., Q-learning.  However, CRL with tuned exploration parameters (e.g., hyperparameters $\epsilon$ and $\tau$) may suffer from difficulty of balancing exploration and exploitation, which can further affect its learning quality and convergence performance.
To give a better alternative for solving problem (1), the QiRL technique will be invoked to tackle the proposed optimal trajectory planning problem.

\vspace{-.0cm}
\pagestyle{empty}
\section{QiRL Solution}
The above trajectory design problem can be interpreted as a sequential decision-making process following Markov property, which means that  the UAV's movement decision for the current time slot can be sorely determined according to the information of the previous time slot, regardless those of time slots before the previous time slot. Therefore, Markov decision process (MDP) is a suitable candidate for solving the trajectory optimization problem, forging the optimal mapping (i.e., the optimal action selection policy) from the state space to the corresponding action selections.
\subsection{The MDP Formulation}
To formulate the MDP, we need to clarify the {\it states} of the proposed QiRL solution for the considered scenario. The feasible area $\Phi$ is divided into $N_{1}$ by $N_{2}$ small grids and the side length of each grid equals $VT$. Besides, we assume that the sum of received signal strength keeps constant within each grid.$\footnote{This assumption is reasonable because the acreage of each grid is far less than that of $\Phi$, in the case of sufficient discretization.}$ The GUs are located in some of the small squares, which will be specified in the numerical results. According to the discrete tabular form of $\Phi$, the state set of the UAV can be written as $\mathcal{S}=\{s_{1},s_{2},\dots,s_{N_{1}N_{2}}\}$, where $s_{i}\in\mathcal{S}$ represents a small square in $\Phi$. Because we focus on the ESUTR maximization problem, it is straightforward to define $R=\sum_{k=1}^{K}\omega_{k}\log(1+\Gamma_{s_{i}k})$ as the {\it reward} function, where $L_{s_{i}}$ in $\Gamma_{s_{i}k}$ denotes the location of a possible state $s_{i}$. In the case of reaching the boundary of $\Psi$, the UAV will be rebounded back and the reward for this trial is set to zero.\footnote{Hereby, we take zero reward for crashing into the boundary as an example. Of course, one can let this kind of scenario be punished by minus reward. } Note that the UAV is only able to observe $R_{s_{i}}$ while other network information is inaccessible, i.e., $P_{k}$, $\omega_{k}$, $\sigma^{2}_{k}$ and $\vec{D}_{k}$. The UAV aims to find an optimal path, in which the ESUTR of the GUs should be the greatest among all possible UAV roaming routes from $\vec{L}_{0}$ to $\vec{L}_{F}$. To drive the UAV to the destination $\vec{L}_{F}$, the UAV will gain a special reward which is defined as $\hat{R} = 10\times\underset{s_{i}\in\mathcal{S}}{\max}R_{s_{i}}$, once it reaches $\vec{L}_{F}$. Regarding the UAV's possible {\it actions}, we limit the movement options of the UAV in the following action set $\mathcal{A}=\{$forward, backward, left, right$\}$, which will be denoted as quantum eigenactions in the proposed QiRL solution. 
The goal of the proposed QiRL algorithm is to learn a mapping from states to actions, i.e., the UAV aims to learn a policy $\pi:\mathcal{S}\times\mathcal{A}\rightarrow[0,1]$ so that the expected sum of discounted rewards for one episode can be maximized. We define the value function (in Bellman equation form) of state $s$ under policy $\pi$ at trial $t$ as
\begin{align}
V_{\pi}\left(s\right) &= \mathbb{E}_{\pi}\left[\sum_{l = 0}^{F}\gamma^{l}R\left(t+l+1\right)\vert s\left(t\right)=s\right]\nonumber\\
&=\mathbb{E}_{\pi}\left[R(t+1)+\gamma\sum_{l = 0}^{F}\gamma^{l}R\left(t+l+2\right)\vert s\left(t\right)=s\right]\nonumber\\
&=\sum_{a\in\mathcal{A}}\pi(s,a)\sum_{s'\in\mathcal{S}}T_{ss'}^{a}\times\nonumber\\
&\left\{R_{ss'}^{a}+\gamma\mathbb{E}_{\pi}\left[\sum_{l=0}^{F}\gamma^{l}R(t+l+2)\vert s(t+1)=s'\right]\right\}\nonumber\\
&=\sum_{a\in\mathcal{A}}\pi(s,a)\sum_{s'\in\mathcal{S}}T^{a}_{ss'}\left[R_{ss'}^{a}+\gamma V_{\pi}\left(s'\right)\right],
\label{Bellman_Value_Function}
\end{align}
where $\gamma\in[0,1]$ represents the discount factor, $R_{ss'}^{a}$ means the immediate reward when reaching $s'$, $\pi(s,a)$ is the probability of picking action $a$ for state $s$ under the policy $\pi$ and $T^{a}_{ss'}=\Pr[s(t+1)=s'\vert s(t)=s, a(t)=a]$ denotes the probability of state transition $s\rightarrow s'$ given action $a$. Note that state transition is deterministic rather than stochastic in the considered model, and thus (\ref{Bellman_Value_Function}) should be rewritten as
\begin{equation}
	V_{\pi}\left(s\right) = \sum_{a\in\mathcal{A}}\pi(s,a)\sum_{s'\in\mathcal{S}}\left[R_{ss'}^{a}+\gamma V_{\pi}\left(s'\right)\right].\label{Value_Function_Rewritten}
\end{equation}
The QiRL algorithm is applied to find the optimal policy $\pi^{*}$. Based on (\ref{Value_Function_Rewritten}), the optimal value function (Bellman optimality equation) can be derived as 
\begin{equation}
V^{*}(s)=\underset{a\in\mathcal{A}}{\max}\sum_{s'\in\mathcal{S}}\left[R_{ss'}^{a}+\gamma
V^{*}(s')\right].\label{Optimal_Value_Function}
\end{equation}
Note that the optimal value function is no-linear and has no closed-form solution, which can be tackled alternatively via iterative methods, e.g., temporal difference (TD) updating \cite{dong2010robust}. Specifically, the TD-based value updating rule of the proposed QiRL can be described as 
\begin{equation}
\label{V_TD_Update}
V\left(s\right)\leftarrow\hspace{.1cm}V\left(s\right)+\alpha\left[R\left(s'\right)+\gamma V\left(s'\right)-V\left(s\right)\right],
\end{equation}
where $s'$ means the next state after taking an action and $\alpha$ indicates the learning rate.
From the above MDP formulation, it is easy to find that the objective function (\ref{ProposedProblem}) corresponds to the un-discounted expected rewards over one episode, which is a special type of MDP called episodic tasks where a special state named the terminal state separates the agent-environment interactions into episodes. Thus, we set the discounted factor $\gamma=1$ for the considered problem. The terminal state and the start state correspond to $\vec{L}_{F}$ and $\vec{L}_{0}$, respectively.

\subsection{Collapsing Action Selection}
According to quantum mechanics \cite{nielsen2010quantum}, a quantum state $\ket{\Psi}$ (Dirac representation) can describe the state of a closed quantum system, which is a unit vector (i.e., $\braket{\Psi} = 1$) in Hilbert space. The quantum state $\ket{\Psi}$ consisting of $n$ quantum bits (qubits) can be expanded as
\begin{equation}
\ket{\Psi} = \ket{\psi_{1}}\otimes\ket{\psi_{2}}\otimes\dots\otimes\ket{\psi_{n}} = \sum_{p = 00...0}^{\overbrace{11...1}^{n}}h_{p}\ket{p},
\end{equation}
where $\ket{\psi_{i}}, i\in\left[1,n\right]$ represents the $i$-th qubit which is a two-state quantum system and the basic unit of quantum information, the complex coefficient $h_{p}$ (subject to $\sum_{p = 00...0}^{11...1}\vert h_{p}\vert^{2} = 1$) denotes the probability amplitude for eigenstate $\ket{p}$ of $\ket{\Psi}$ and $\otimes$ represents the tensor product. The representation of $n$-qubit quantum state $\ket{\Psi}$ follows the quantum phenomenon called \textit{state superposition principle}. 
Note that $h_{p}$ can take $2^{n}$ complex values so that the $n$-qubit quantum state $\ket{\Psi}$ can be regarded as the superposition of $2^{n}$ eigenstates, in the range from $\ket{00...0}$ to $\ket{11...1}$. 

To represent the four possible actions in QiRL, two qubits are sufficient. Furthermore, eigenactions (i.e., the quantum representation of physical actions) $\ket{a_{1}},\ket{a_{2}},\ket{a_{3}},\ket{a_{4}}$ are allocated to denote the actions forward, backward, left and right, respectively. Inspired by the superposition principle of quantum theory, we can represent the four egienactions in their quantum superposition form, given by
\begin{equation}
\ket{A(l)} = \ket{\psi_{1}}\otimes\ket{\psi_{2}}=\sum_{a = 00}^{11}h_{a}\ket{a}\rightarrow\sum_{n = 1}^{4}h_{n}\ket{a_{n}},
\end{equation}
where $l$ represents a specific trial and the complex coefficients $h_{n}$ and $h_{a}$ are the probability amplitudes under the normalisation constraints $\sum_{n=1}^{4}\vert h_{n}\vert^{2}\!=\!1$ and $\sum_{a = 00}^{11}\vert h_{a}\vert^{2}\!=\!1$, respectively. Note that the two-qubit superposition $\ket{A(l)}$ is a unit vector in a 4-dimensional Hilbert space spanned by the four orthogonal bases $\ket{00}$, $\ket{01}$, $\ket{10}$ and $\ket{11}$. Specifically, the action taken by the UAV before any quantum measurement lies in a superposition state (four options in total, i.e., $\ket{a_{1}}$, $\ket{a_{2}}$, $\ket{a_{3}}$ and $\ket{a_{4}}$), which is mapped into the tensor product of two qubits. 

In quantum theory, when an external agency (e.g., observer, experimenter) measures the quantum state $\ket{\Psi} = \sum_{n}\varrho_{n}\ket{\psi_{n}}$ with the eigenbasis $\{\psi_{n}\}$, $\ket{\Psi}$ will collapse from the superposition state to one of its eigenstates $\ket{\psi_{n}}$, i.e., $\ket{\Psi} \rightarrow \ket{\psi_{n}}$, with probability $\vert\bra{\psi_{n}}\ket{\Psi}\vert^{2} = \vert\varrho_{n}\vert^{2}$. Inspired by this \textit{quantum collapse phenomenon}, the superposition $\ket{A(l)}$ is supposed to collapse onto one of the eigenactions $\ket{a_{n}}$ with probability of $\vert h_{n}\vert^{2}$, during the action picking period in the proposed QiRL algorithm.
\subsection{Grover Iteration}
The quantum representation $\ket{A(l)}$ establishes a bridge between quantum eigenactions and the physical action set $\mathcal{A}$, which allows us to apply quantum amplitude amplification as a reinforcement strategy. The probability amplitude of each eigenaction can be amplified or attenuated via specific quantum algorithm (e.g., Grover's iteration\cite{nielsen2010quantum}), gradually modifying the probability distribution of collapsing. To realize this, two unitary operators can be employed for the currently chosen action $\ket{a_{i}}$ which is from the $l$-th trial $\ket{A(l)}= \sum_{n = 1}^{4}h_{n}\ket{a_{n}} = h_{i}\ket{a_{i}}+h_{a_{i}^{\perp}}\ket{a_{i}^{\perp}}$, shown as
\begin{equation}
\boldsymbol{U}_{\ket{a_{i}}} = \boldsymbol{I} - (1 - e^{j\phi_{1}})\ket{a_{i}}\bra{a_{i}},
\end{equation}
\begin{equation}
\boldsymbol{U}_{\ket{A(l)}} = (1 - e^{j\phi_{2}})\ket{A(l)}\bra{A(l)} - \boldsymbol{I},
\end{equation}
where $\ket{a_{i}^{\perp}}=\sum_{n\neq i}\frac{h_{n}}{h_{a_{i}^{\perp}}}\ket{a_{n}}$ means the vector orthogonal to $\ket{a_{i}}$, $h_{a_{i}^{\perp}} = \sqrt{\sum_{n\neq i}\vert h_{n}\vert^{2}} = \sqrt{1 - \vert h_{i}\vert^{2}}$, $\boldsymbol{I}$ represents the identity matrix and $\bra{a_{n}}$ and $\bra{A(l)}$ are Hermitian transposes of $\ket{a_{n}}$ and $\ket{A(l)}$, respectively. Then, the Grover operator can be constructed in the form of unitary transformation, given by
$\boldsymbol{G} = \boldsymbol{U}_{\ket{A(l)}}\boldsymbol{U}_{\ket{a_{i}}}.$
After $m$ times of applying $\boldsymbol{G}$ on $\ket{A(l)}$, the amplitude vector in the next trial becomes
$\ket{A(l+1)} = \boldsymbol{G}^{m}\ket{A(l)}.$

There are mainly two methods to deal with the aforementioned probability amplitude updating task. One is to choose a feasible value of $m$ with fixed parameters $\phi_{1}$ and $\phi_{2}$ (commonly both of them equal to $\pi$); the other is to fix $m=1$ with dynamic parameters $\phi_{1}$ and $\phi_{2}$. Because the former updating approach can only modify the amplitudes in a discrete manner, the later method is chosen in this work, i.e., Grover iteration with flexible parameters $\phi_{1}$ and $\phi_{2}$. Then, the impacts of $\boldsymbol{G}$ on the superposition representation $\ket{A(l)}$ can be given by the following proposition.
\begin{prop}
	\label{thmGonAl}
	The overall effects of $\boldsymbol{G}$ with free parameters $\phi_{1}$ and $\phi_{2}$ on the superposition representation $\ket{A(l)}$ at the $l$-th trial can be expressed analytically as
	\begin{equation}
	\boldsymbol{G}\ket{A(l)}=(\mathcal{Q}-e^{j\phi_{1}})h_{i}\ket{a_{i}} + (\mathcal{Q} - 1) h_{a_{i}^{\perp}}\ket{a_{i}^{\perp}},
	\end{equation}
where $\mathcal{Q} = (1 - e^{j\phi_{2}})\left[1 - (1 - e^{j\phi_{1}})\vert h_{i}\vert^2\right]$.
\end{prop}
\begin{IEEEproof}
The impacts of $\boldsymbol{U}_{\ket{a_{i}}}$ on $\ket{a_{i}}$ and $\ket{a_{i}^{\perp}}$ can be given by\vspace{-.1cm}
\begin{equation}
\boldsymbol{U}_{\ket{a_{i}}}\ket{a_{i}}=\left[ \boldsymbol{I} - (1 - e^{j\phi_{1}})\ket{a_{i}}\bra{a_{i}}\right]\ket{a_{i}}=e^{j\phi_{1}}\ket{a_{i}},\vspace{-.1cm}
\end{equation} 
\begin{equation}
\boldsymbol{U}_{\ket{a_{i}}}\ket{a_{i}^{\perp}}=\left[ \boldsymbol{I} - (1 - e^{j\phi_{1}})\ket{a_{i}}\bra{a_{i}}\right]\ket{a_{i}^{\perp}}=\ket{a_{i}^{\perp}},
\end{equation} respectively.
Furthermore, we have 
\begin{align}
\boldsymbol{U}_{\ket{a_{i}}}\ket{A(l)}&=\left[ \boldsymbol{I} - (1 - e^{j\phi_{1}})\ket{a_{i}}\bra{a_{i}}\right]\ket{A(l)} \nonumber \\
&=e^{j\phi_{1}}h_{i}\ket{a_{i}} + h_{a_{i}^{\perp}}\ket{a_{i}^{\perp}},
\end{align} 
in which $\boldsymbol{U}_{\ket{a_{i}}}$ plays the role as a conditional phase shift operator in quantum computation. At the end, we can obtain
\begin{align}
&\boldsymbol{G}\ket{A(l)}=\boldsymbol{U}_{\ket{A(l)}}\boldsymbol{U}_{\ket{a_{i}}}\ket{A(l)} \nonumber \\
&=(1 - e^{j\phi_{2}})\left[h_{i}\ket{a_{i}}+h_{a_{i}^{\perp}}\ket{a_{i}^{\perp}}\right]\times \nonumber \\
&\left[h_{i}^{\dagger}\bra{a_{i}}+h_{a_{i}^{\perp}}^{\dagger}\bra{a_{i}^{\perp}}\right]\boldsymbol{U}_{\ket{a_{i}}}\ket{A(l)} - \boldsymbol{U}_{\ket{a_{i}}}\ket{A(l)} \nonumber \\
&=(\mathcal{Q}-e^{j\phi_{1}})h_{i}\ket{a_{i}} + (\mathcal{Q} - 1) h_{a_{i}^{\perp}}\ket{a_{i}^{\perp}},
\end{align} 
where $\mathcal{Q} =(1 - e^{j\phi_{2}})\left[1 - (1 - e^{j\phi_{1}})\vert h_{i}\vert^2\right].$
\end{IEEEproof}

\begin{remrk}
	The ratio between the probability amplitudes of $\ket{a_{i}}$ after being acted by the Grover operator $\boldsymbol{G}$ and before that can be expressed as
	\begin{align}
		\mathcal{R} 
		&=(1 - e^{j\phi_{1}} - e^{j\phi_{2}}) - (1 - e^{j\phi_{1}})(1 - e^{j\phi_{2}})\vert h_{i}\vert^2.
	\end{align} 
		Then, the updated occurrence probability of the selected action $\ket{a_{i}}$ after the Grover iteration can be given by $\vert\mathcal{R}\vert^2\vert h_{i}\vert^2$.
\end{remrk}
\begin{remrk}
	For ease of understanding the effect of $\boldsymbol{G}$, we show its corresponding algebraic visualization. In Fig. 2, $\ket{A(l)}$ is reconstructed in the form of polar coordinates on the Bloch sphere, shown as
	\begin{align}
		\ket{A(l)} 
		&=e^{j\zeta}(\cos\frac{\theta}{2}\ket{a_{i}} + e^{j\varphi}\sin{\frac{\theta}{2}}\ket{a_{i}^{\perp}}) \nonumber \\
		&\simeq \cos\frac{\theta}{2}\ket{a_{i}} + e^{j\varphi}\sin{\frac{\theta}{2}}\ket{a_{i}^{\perp}},
	\end{align}
	where the parameter $e^{j\zeta}$ can be omitted, since a global phase poses no observable effects \cite{li2020quantum}. Note that the polar angle parameter $\theta$ and the azimuthal angle variable $\varphi$ define the unit vector $\ket{A(l)}$ on the Bloch sphere, as shown in Fig. 2. 
	The impact of $\boldsymbol{U}_{\ket{a_{i}}}$ can be understood as a clockwise rotation around the $z$-axis by $\phi_{1}$ (the red circle) on the Bloch sphere, leading to the rotation from $\ket{A(l)}$ to $\ket{A(l)'}$. Similarly, if we change the basis from $\{\ket{a_{i}},\ket{a_{i}^{\perp}}\}$ to $\{\ket{A(l)}, \ket{A(l)^{\perp}}\}$, $\boldsymbol{U}_{\ket{A(l)}}$ makes a clockwise rotation around the new $z$-axis $\ket{A(l)}$ by $\phi_{2}$ (the blue circle), which rotates $\ket{A(l)'}$ to $\ket{A(l+1)}$. Therefore, the overall effect of $\boldsymbol{G}$ on $\ket{A(l)}$ is a two-step rotation which can modify the polar angle $\theta$, when the basis is locked as $\{\ket{a_{i}},\ket{a_{i}^{\perp}}\}$. Via controlling parameters $\phi_{1}$ and $\phi_{2}$, it is possible to realize arbitrary parametric rotation on the Bloch sphere, which acts as the foundation for modifying the probability amplitudes of $\ket{A(l)}$. The smaller $\theta$ is, the higher probability $\ket{A(l)}$ will collapse to $\ket{a_{i}}$ when it is measured, which inspires us to apply it as a reinforcement strategy. The core of this reinforcement approach is to achieve a smaller $\theta$ via manipulating $\phi_{1}$ and $\phi_{2}$ when $\ket{a_{i}}$ is recognized as a "good" action. Otherwise, if $\ket{a_{i}}$ is determined as a "bad" action, $\phi_{1}$ and $\phi_{2}$ should be modified to enlarge $\theta$.
	
	\begin{figure}
		\centering
		\includegraphics[width=0.25\textwidth]{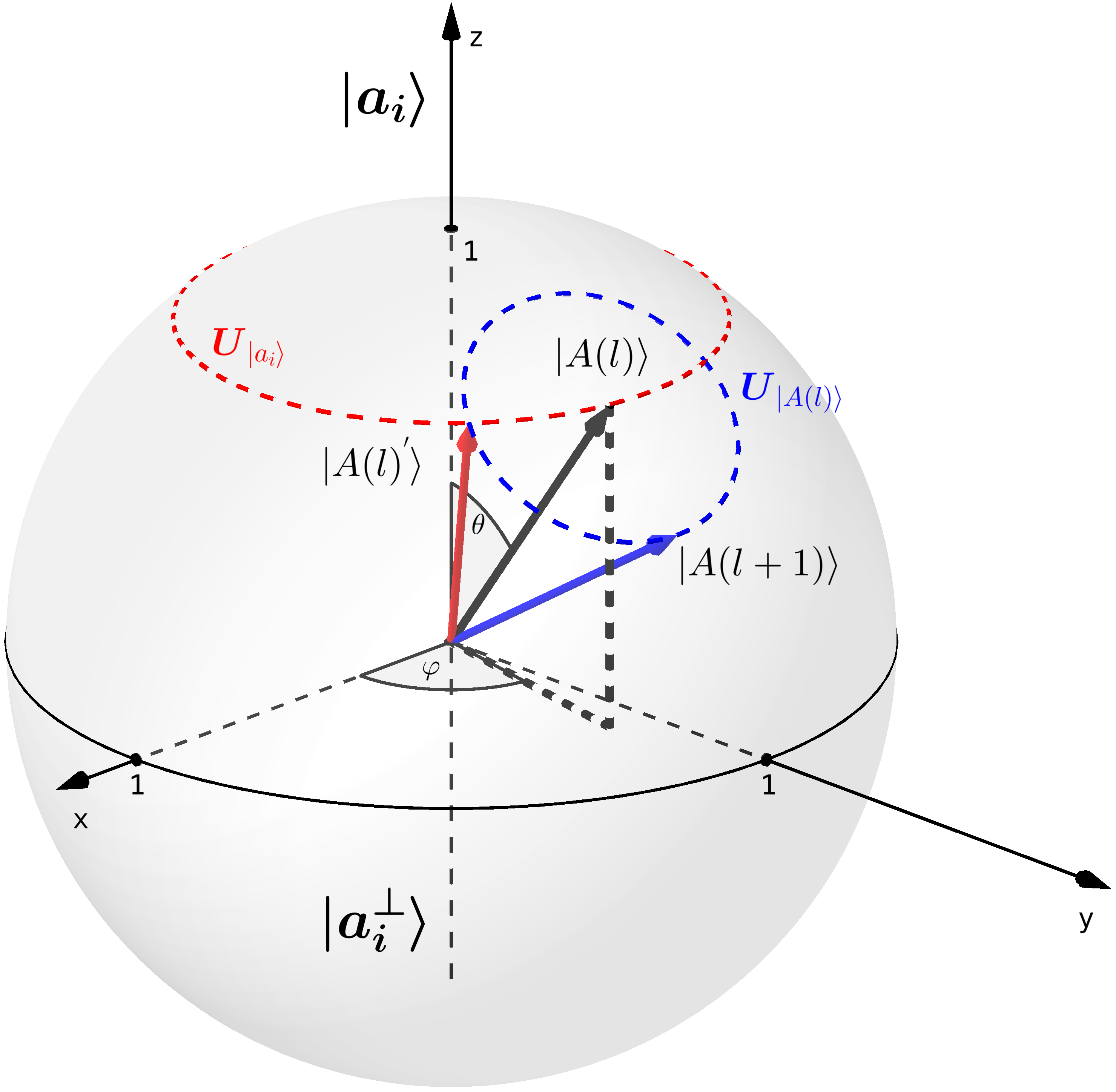}
		\caption{Geometric explanation of the Grover rotation}
		\label{Fig_GeometricExplation_Grover3D}
		\vspace{-.5cm}
	\end{figure}
\end{remrk}

\subsection{The Proposed QiRL Algorithm}

Remark 1 and Remark 2 give an explanation for amplitude amplification in quantum mechanism, which can be applied as the quantum-inspired reinforcement strategy for our proposed QiRL approach. According to Remark 1, it is straightforward to conclude that $\vert\mathcal{R}\vert^2$ should be designed to be larger than 1, if the current representation $\ket{a_{i}}$ is determined as a "good" action. Otherwise, $\vert\mathcal{R}\vert^2$ should be manipulated to be smaller than 1. By selecting feasible $\phi_{1}$ and $\phi_{2}$, it is possible to manipulate the value of $\vert\mathcal{R}\vert^2$ in the manner as mentioned before, which can be interpreted geometrically via Remark 2. For the sake to simulate it on the conventional computer, we use $e^{k*[R+V(s')]}$ to represent alternatively the overall effects of $\boldsymbol{G}$ on probability $\vert h_{i} \vert^2$, which means the updated occurrence probability of the selected action $\ket{a_{i}}$ should be $e^{k*[R+V(s')]}\vert h_{i} \vert^2$. If $k>0$, the current action will be rewarded while it will be punished if $k<0$. The updating amplification is controlled via $k*[R+V(s')]$.


Note that all the possible probability amplitudes together should be re-normalized after each implementation of amplitude amplification, which is subject to the normalization constraint of quantum superposition. The proposed QiRL solution is concluded in Algorithm 1, which can be conducted in conventional computers.

\begin{remrk}
 The quantum-inspired reinforcement strategy prioritizes all possible actions in ranked probability sequence which is gradually updated alongside the learning process. Thus, it can naturally balance the exploration and exploitation, in which no tuned exploration parameter is necessary. This enhancement has the potential to help realize faster convergence and satisfactory learning quality, which will be later illustrated in the simulation results.
\end{remrk}

\begin{prop}
	The convergence of the proposed QiRL algorithm is guaranteed when the learning rate $\alpha$ is non-negative and satisfies $\underset{T\rightarrow\infty}{\lim}\sum_{k=1}^{T}\alpha_{k}=\infty$ and $\underset{T\rightarrow\infty}{\lim}\sum_{k=1}^{T}\alpha_{k}^{2}<\infty$.
\end{prop}

\begin{IEEEproof}
The proof is omitted for its simplicity, which is similar to the proof of Proposition 2 in \cite{dong2008quantum}.
\end{IEEEproof}

\begin{algorithm}
	\SetKwData{Ini}{\bf{Initialization:}}
	\caption{The proposed QiRL algorithm}\label{CRLalgorithm}
	\KwIn{Learning parameters: $\alpha\in\left[0,1\right]$, $\gamma=1$; UAV informations:
		$\vec{L}_{0}$, $\vec{L}_{F}$, $H$, $V$, $T$, $E$;}
	\KwOut{The optimal policy $\pi^{*}$=AmpMem;}
	\Ini $ep = 0$; s = $\vec{L}_{0}$; $V\left(s\right) = 0$, $\forall$ $s\in\mathcal{S}$; AmpMem = defaultdict($lambda$: [$\frac{1}{4}$, $\frac{1}{4}$, $\frac{1}{4}$, $\frac{1}{4}$]);
	
		\While{ep $\leq$ NumEp} {
		\Repeat{$F> E/T$ \rm{or} $s'==\vec{L}_{F}$}{Choose a feasible $a$ for $s$ via measuring AmpMem[s]\;	
			Apply $a$, then observe the reward $R$ and the next state $s'$\;
			Update the value function as per\\
			$V\left(s\right)\leftarrow\hspace{.1cm}V\left(s\right)+\alpha\left[R+\gamma V\left(s'\right)-V\left(s\right)\right]$\;
			Apply quantum-inspired reinforcement factor $e^{k*[R+V(s')]}$ on AmpMem[s][a]. When the UAV hits the boundary or $\Delta V(s)<0$, $k<0$. Otherwise, $k>0$\;
			Re-normalize AmpMem[s]\;
				$s\leftarrow s'$		}
		$ep$ += 1\;
	}
\end{algorithm}
\vspace{-.5cm}
\section{Simulation Results}
\begin{figure*}[htbp] 
	\centering  
	\subfigure[Accumulated Reward Comparison]{
		\label{level.sub.1}
		\includegraphics[width=.32\linewidth]{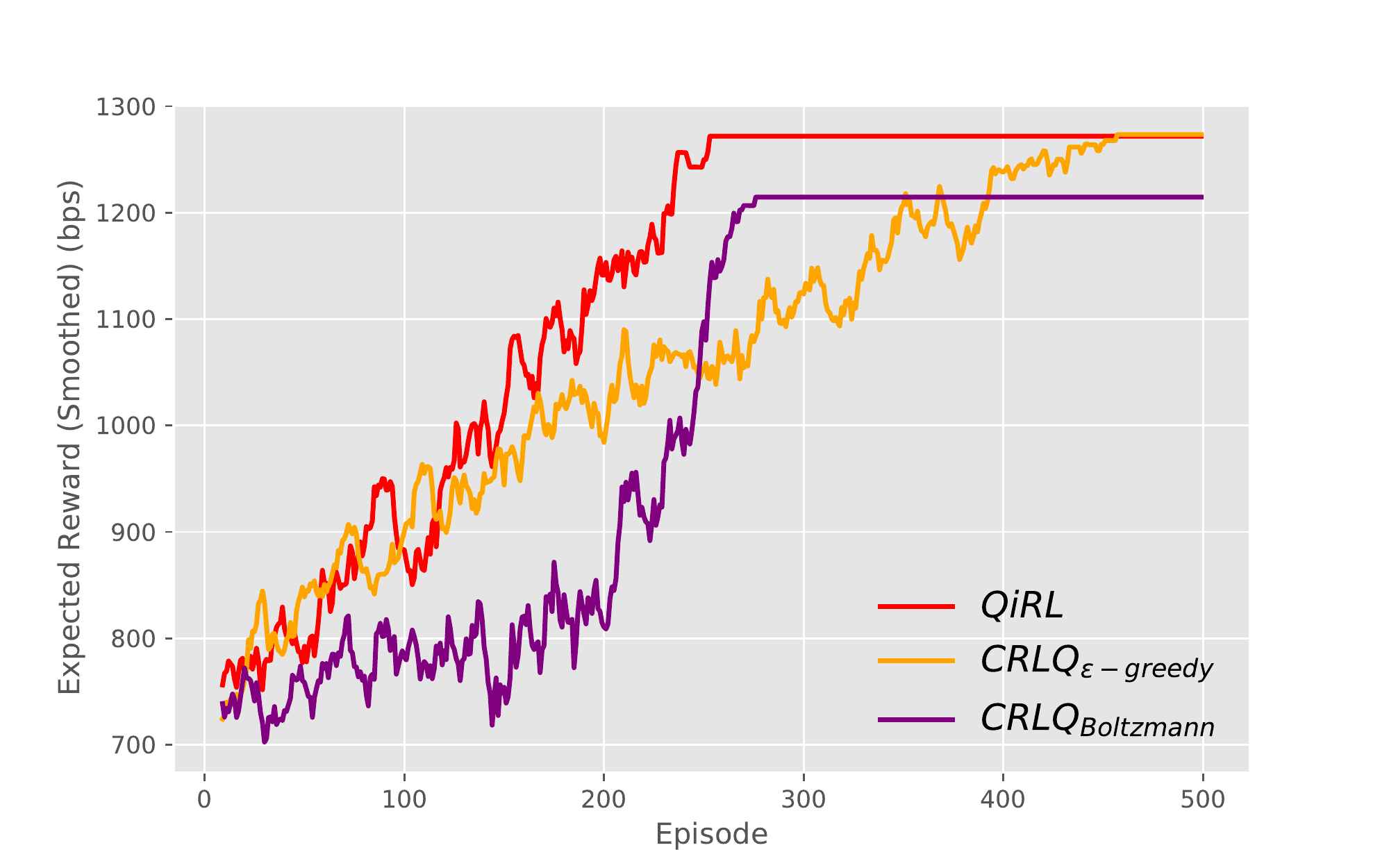}}
	\subfigure[Learned Trajectory Comparison (Env. 1)]{
		\label{level.sub.2}
		\includegraphics[width=.32\linewidth]{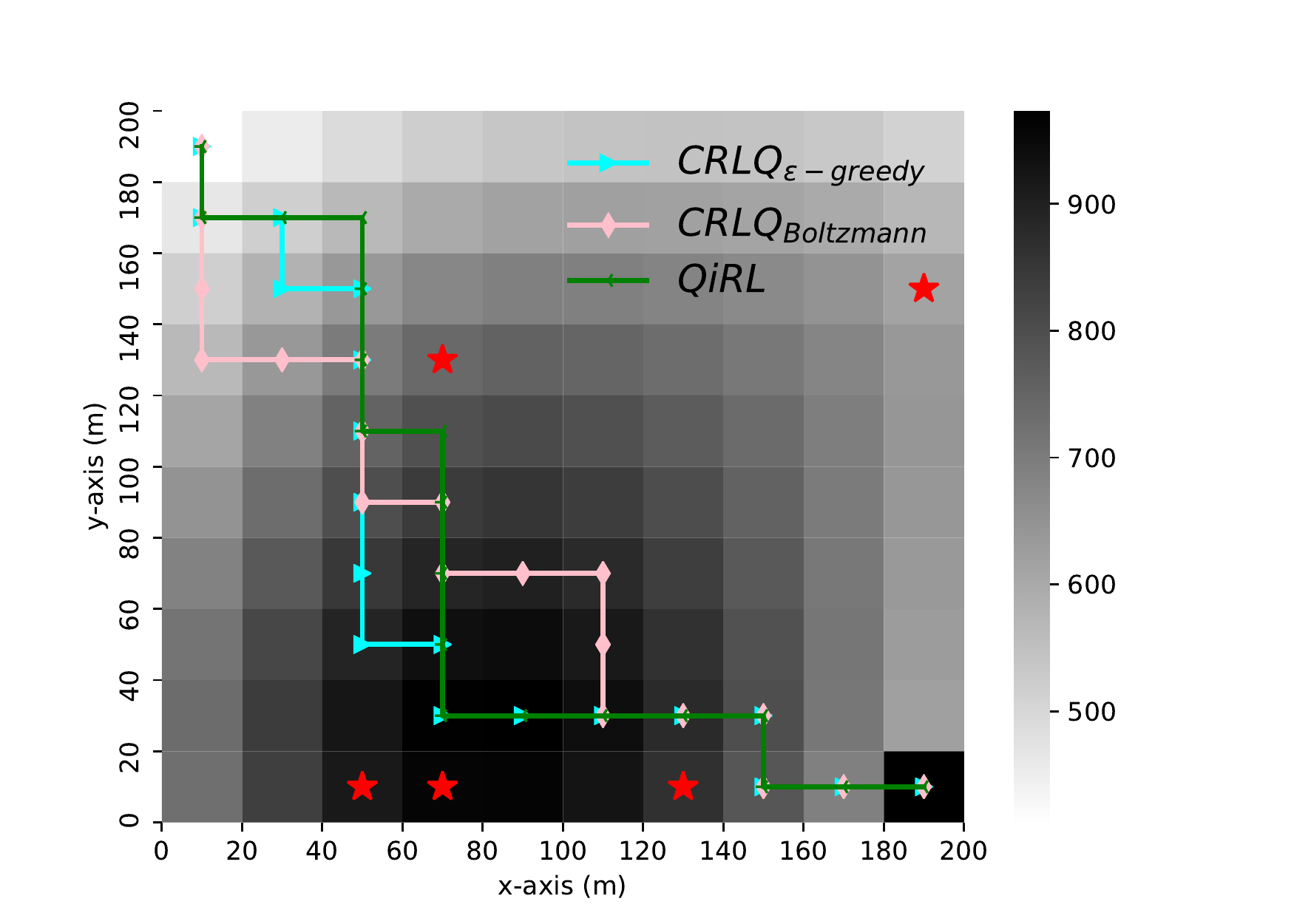}}
	\subfigure[Learned Trajectory Comparison (Env. 2)]{
		\label{level.sub.3}
		\includegraphics[width=.32\linewidth]{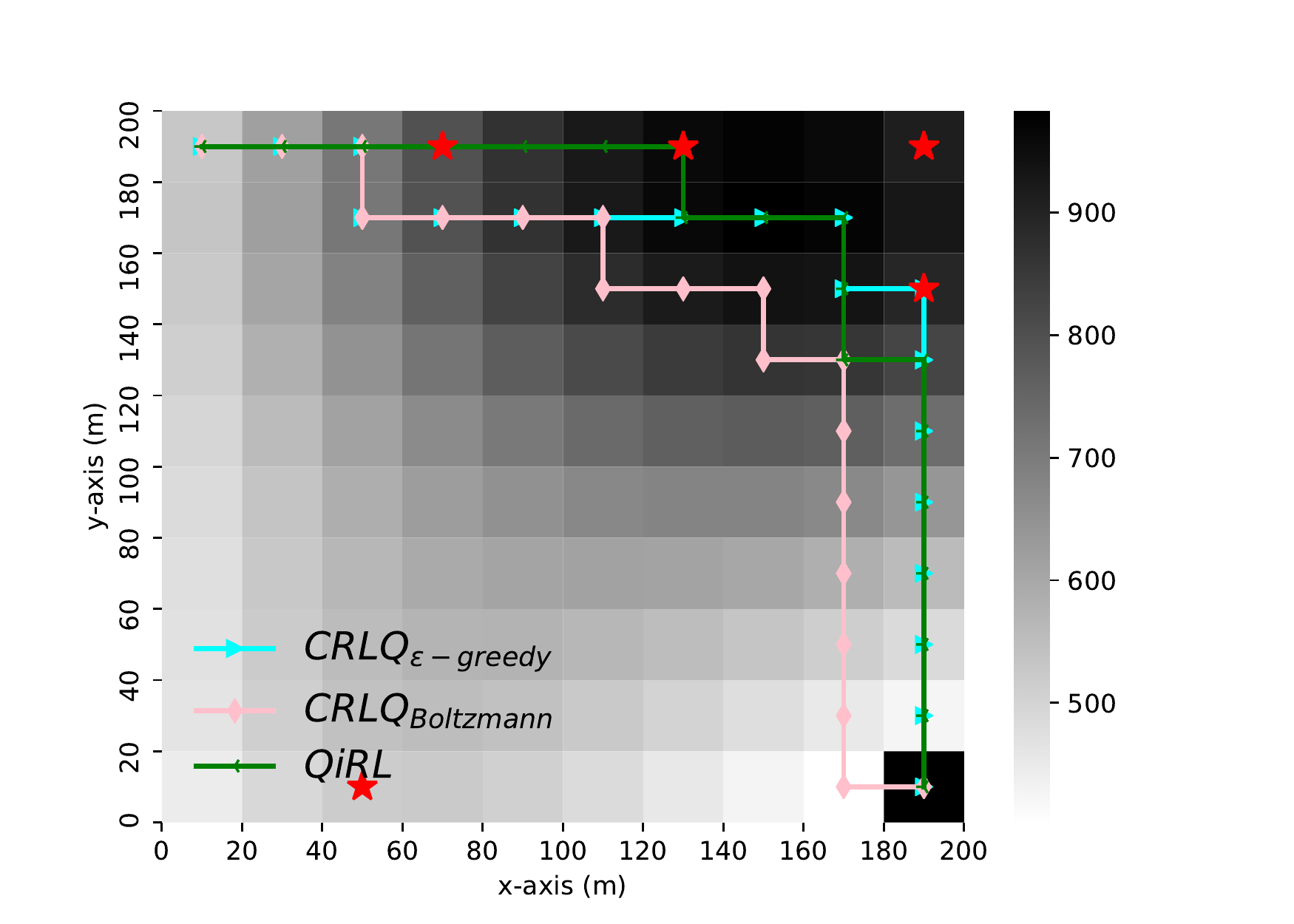}}\\

\vspace{-.2cm}
    \caption{Performance comparison of two Q-learning approaches and the proposed QiRL solution}
	\label{level}
	\vspace{-.5cm}
\end{figure*}
In this section, experimental results are evaluated for the considered UAV trajectory planning problem via the proposed QiRL solution. For comparison, two CRL methods (i.e., Q-learning with $\epsilon$-greedy and Boltzmann exploration strategies) are also performed as benchmarks. It is assumed that the feasible UAV exploration field $\Phi$ is a square area with side length $200$ m, where 5 GUs are located on the ground (denoted by the red stars).  By default, the length of each time slot is fixed as $T\!=\!2$ s and the constant flying altitude and speed of the UAV are set as $H\!=\!100$ m and $V\!=\!10$ m/s, respectively. The area $\Phi$ is divided into $10$-by-$10$ small grids and the side length of each grid equals $VT\!=\!20$ m. The start location and the destination are predefined at $\vec{L}_{0}=\left(10,190,100\right)$ and $\vec{L}_{F}=\left(190,10,100\right)$, respectively. Considering the on-board power capacity of the UAV, the total flying time of the UAV is constrained as $FT \leq 1800$ s so that we set $E=1800$. Besides, we set $P_{k}=1$ Watt, $\sigma_{k}^{2}=1$, $\varphi=2$ GHz, $B=10$ MHz and $\omega_{k}=2$ MHz, which is in line with \cite{yin2019intelligent}.

Fig. 3 shows the performance comparison of one widely-used CRL approach called Q-learning with two action selection strategies, i.e., $\epsilon$-greedy and Boltzmann, and the proposed QiRL solution. Note that exploration parameters $\epsilon$ and $\tau$ of Q-learning approach keep annealing alongside the learning progress, which controls the ratio of exploration and exploitation and highly affects the overall learning quality and convergence performance of Q-learning approach. In this figure, the learned trajectories of Q-learning and QiRL are also depicted for intuitive comparison. Specifically, subfigure (a) shows the expected reward curves, which corresponds to subfigure (b).

 From subfigure (a), it is straightforward to observe that the proposed QiRL solution can converge much faster than Q-learning with $\epsilon$-greedy action selection strategy, while it has relatively faster convergence speed than Q-learning with advanced Boltzmann action selection strategy, which illustrates that the proposed QiRL algorithm can offer better convergence performance. Moreover, from subfigures (b) and (c), we can observe that all the simulated RL approaches can output proper trajectories in these two different network environments. However, while Boltzmann strategy can offer faster convergence performance than $\epsilon$-greedy, it leads to sub-optimal trajectory, as shown in subfigures (a) and (b). According to Fig. 3, the proposed QiRL solution can not only enhance convergence performance but also achieve the equivalently optimal trajectory compared to Q-learning with $\epsilon$-greedy action selection strategy. Note that the balancing between exploration and exploitation in $\epsilon$-greedy or Boltzmann aided Q-learning approach is controlled by the pickings of initial exploration parameter (i.e., $\epsilon$ or $\tau$, respectively) and their corresponding annealing speeds, which directly and inherently influences convergence performance and learning quality. Generally speaking, the initial exploration parameters and their corresponding annealing speeds are modified via empirical knowledge when the learning environment varies. However, simply decaying exploration parameter (linearly or non-linearly) alongside the learning progress could easily lead to insufficient learning or low speed of convergence. To deal with this unsatisfactoriness, the proposed QiRL algorithm applies quantum-inspired action selection approach, offering natural balancing between exploration and exploitation alongside the learning progress and thus can better deal with the trade-off between convergence speed and learning quality.

\section{Conclusion}
This paper introduced a QiRL solution to tackle the trajectory planning problem which aims to maximize the ESUTR performance for the UAV flying from the start location to the destination. Specifically, the proposed QiRL approach utilizes the novel collapse action selection strategy inspired by quantum mechanism, which can offer a natural way to balance exploration and exploitation via sorting probabilities of action collapse in a ranking sequence. Numerical results compared the convergence performance and the learned trajectories between the proposed QiRL solution and the widely-used Q-learning approach with $\epsilon$-greedy and Boltzmann exploration strategies, validated the effectiveness of the proposed QiRL solution and showed that the QiRL solution can better deal with the trade-off between convergence speed and learning quality than traditional Q-learning approaches.
\bibliographystyle{IEEEtran}
\bibliography{reference-icc}

\begin{thebibliography}{10}
\providecommand{\url}[1]{#1}
\csname url@samestyle\endcsname
\providecommand{\newblock}{\relax}
\providecommand{\bibinfo}[2]{#2}
\providecommand{\BIBentrySTDinterwordspacing}{\spaceskip=0pt\relax}
\providecommand{\BIBentryALTinterwordstretchfactor}{4}
\providecommand{\BIBentryALTinterwordspacing}{\spaceskip=\fontdimen2\font plus
\BIBentryALTinterwordstretchfactor\fontdimen3\font minus
  \fontdimen4\font\relax}
\providecommand{\BIBforeignlanguage}[2]{{%
\expandafter\ifx\csname l@#1\endcsname\relax
\typeout{** WARNING: IEEEtran.bst: No hyphenation pattern has been}%
\typeout{** loaded for the language `#1'. Using the pattern for}%
\typeout{** the default language instead.}%
\else
\language=\csname l@#1\endcsname
\fi
#2}}
\providecommand{\BIBdecl}{\relax}
\BIBdecl

\bibitem{zeng2019accessing}
Y.~Zeng, Q.~Wu, and R.~Zhang, ``Accessing from the sky: A tutorial on {UAV}
  communications for {5G} and beyond,'' \emph{Proceedings of the IEEE}, vol.
  107, no.~12, pp. 2327--2375, 2019.

\bibitem{wang2017taking}
J.~Wang, C.~Jiang, Z.~Han, Y.~Ren, R.~G. Maunder, and L.~Hanzo, ``Taking drones
  to the next level: Cooperative distributed unmanned-aerial-vehicular networks
  for small and mini drones,'' \emph{IEEE Veh. Technol. mag.}, vol.~12, no.~3,
  pp. 73--82, 2017.

\bibitem{hu2020reinforcement}
J.~Hu, H.~Zhang, L.~Song, Z.~Han, and H.~V. Poor, ``Reinforcement learning for
  a cellular internet of {UAV}s: protocol design, trajectory control, and
  resource management,'' \emph{IEEE Wireless Commun.}, vol.~27, no.~1, pp.
  116--123, 2020.

\bibitem{yin2019intelligent}
S.~Yin, S.~Zhao, Y.~Zhao, and F.~R. Yu, ``Intelligent trajectory design in
  {UAV}-aided communications with reinforcement learning,'' \emph{IEEE Trans.
  Veh. Technol.}, vol.~68, no.~8, pp. 8227--8231, 2019.

\bibitem{challita2019interference}
U.~Challita, W.~Saad, and C.~Bettstetter, ``Interference management for
  cellular-connected {UAV}s: A deep reinforcement learning approach,''
  \emph{IEEE Trans. Wireless Commun.}, vol.~18, no.~4, pp. 2125--2140, 2019.

\bibitem{dong2008quantum}
D.~Dong, C.~Chen, H.~Li, and T.-J. Tarn, ``Quantum reinforcement learning,''
  \emph{IEEE Trans. Syst. Man. Cybern. B}, vol.~38, no.~5, pp. 1207--1220,
  2008.

\bibitem{wang2020thirty}
J.~Wang, C.~Jiang, H.~Zhang, Y.~Ren, K.-C. Chen, and L.~Hanzo, ``Thirty years
  of machine learning: The road to pareto-optimal wireless networks,''
  \emph{IEEE Commun. Surveys Tuts.}, vol.~22, no.~3, pp. 1472--1514, 2020.

\bibitem{li2020quantum}
J.-A. Li, D.~Dong, Z.~Wei, Y.~Liu, Y.~Pan, F.~Nori, and X.~Zhang, ``Quantum
  reinforcement learning during human decision-making,'' \emph{Nature Human
  Behaviour}, vol.~4, no.~3, pp. 294--307, 2020.

\bibitem{biamonte2017quantum}
J.~Biamonte, P.~Wittek, N.~Pancotti, P.~Rebentrost, N.~Wiebe, and S.~Lloyd,
  ``Quantum machine learning,'' \emph{Nature}, vol. 549, no. 7671, pp.
  195--202, 2017.

\bibitem{dong2010robust}
D.~Dong, C.~Chen, J.~Chu, and T.-J. Tarn, ``Robust quantum-inspired
  reinforcement learning for robot navigation,'' \emph{IEEE/ASME Trans.
  Mechatronics}, vol.~17, no.~1, pp. 86--97, 2010.

\bibitem{fakhari2013quantum}
P.~Fakhari, K.~Rajagopal, S.~Balakrishnan, and J.~Busemeyer, ``Quantum inspired
  reinforcement learning in changing environment,'' \emph{New Mathematics and
  Natural Computation}, vol.~9, no.~03, pp. 273--294, 2013.

\bibitem{nielsen2010quantum}
M.~A. Nielsen and I.~L. Chuang, \emph{Quantum computation and quantum
  information}.\hskip 1em plus 0.5em minus 0.4em\relax Cambridge University
  Press, 2010.

\end{thebibliography}

\end{document}